\documentclass[preprint,12pt]{elsarticle}

\usepackage{graphicx}
\usepackage{hyperref}
\usepackage{color}
\hypersetup{
  colorlinks   = true,	%Colours links instead of ugly boxes
  urlcolor     = blue,	%Colour for external hyperlinks
  linkcolor    = blue,	%Colour of internal links
  citecolor   = blue	%Colour of citations
}
\usepackage{amssymb}
 \usepackage{amsthm}
 \usepackage{lipsum,widetext}
\usepackage[T1]{fontenc}
\begin{document}

\begin{frontmatter}

\title{Genesis of chimera patterns through self-induced stochastic resonance}

\author[1]{Taniya Khatun}
\author[1]{Tanmoy Banerjee \corref{cor1}}
\ead{tbanerjee@phys.buruniv.ac.in}
\cortext[cor1]{Corresponding author}
\address[1]{Chaos and Complex Systems Research Laboratory, Department of Physics, University of Burdwan, Burdwan 713 104, West Begal, India}

\begin{abstract}
Noise induced order in excitable systems has diverse manifestations, such as coherence resonance (CR) and stochastic resonance. In this context a less explored phenomenon is self-induced stochastic resonance (SISR). Unlike CR, SISR may arise away from the bifurcation threshold and the properties of the induced oscillations depend upon both the noise intensity and the time-scale separation factor.
In this work, we report a new chimera pattern in a network of coupled excitable units, namely the self-induced stochastic resonance chimera or SISR-chimera that originates from the SISR phenomenon. We explore the detailed dynamics of the SISR-chimera in the parameter space using proper quantitative measures. We have found that unlike CR chimera, the SISR-chimera pattern strongly depends upon the ratio of time scale and noise intensity. Therefore, this type of chimera pattern can be induced even for a tiny noise intensity if the time scale separation of the activator and inhibitor is large enough.
\end{abstract}


%\item A new chimera pattern that originates from self-induced stochastic resonance is discovered.
%\item The self-induced stochastic resonance chimera state is characterized.
%\item The study bridges the gap between chimeras and noise induced order.


\end{frontmatter}


\section{Introduction}
Noise and fluctuations are inevitable in natural processes and systems. The role of noise becomes more prominent and sometimes counterintuitive in excitable systems \cite{vadim-book}. For example, the interplay of noise and excitability gives rise to noise-induced order in the form of coherence resonance (CR) \cite{fhn48} and stochastic resonance (SR) \cite{SR} phenomena. Both of these phenomena are in the center of research for the last few decades \cite{wiesenfeld1995stochastic,Pisarchik2023,lindner2004effects,anna-multiplex,Yamakou2021}. In this context, a less explored phenomenon is the self-induced stochastic resonance reported in Ref.~\cite{sisr,sisr21}. Ref.~\cite{sisr} established that SISR is indeed a different phenomenon than CR: in particular CR arises only near the Hopf bifurcation and is rather insensitive against the noise intensity and the time-scale separation factor. On the other hand, SISR may arise away from the bifurcation threshold and the properties of the induced oscillations depend upon both the noise intensity and the time-scale separation factor \cite{sisr-fred}.
%%
%Ref.~\cite{sisr} showed that, in contrast to the SISR, in the case of CR, the statistics of time period of limit cycle are largely independent of the time scale separation when the noise is added only to activator and not to the inhibitor. 

In nature, dynamical systems are rarely isolated and, therefore, coupled oscillators offer a general framework to study the effect of coupling or interaction among the dynamical units. Several emergent dynamics have been studied in this context, e.g., synchronization, pattern formation, and symmetry-breaking \cite{ZAK20}. The most prominent emergent behavior that has been identified lately is the chimera pattern, where synchrony-asynchrony patterns coexist as a result of symmetry breaking in a network of identical oscillators \cite{ZAK20}. Since its discovery in Ref.~\cite{kuramoto}, several chimera patterns have been discovered in theory \cite{amc_sethia,Larger_2013,schoell-CD,tan-CD,schoell_qm,tbpre,kemeth,csod,lr16,Semenov2016} and experiments \cite{raj,keneth} (see \cite{ZAK20} and \cite{chireview,diba,schoell_rev,chirev-rep} for detail). 
However, the interplay of chimera and noise is relatively a less understood topic in the study of chimeras \cite{Loos_2016}.   
Only recently, Ref.~\cite{cr-chimera} reported a novel chimera pattern, namely the CR chimera in coupled excitable FitzHugh-Nagumo units near the Hopf bifurcation threshold. Later, Ref.~\cite{type1} generalizes the notion of CR chimera for type-I excitable systems. Ref.~\cite{acanna} proposed the control scheme through time delay of CR chimera patterns. These studies motivate us to ask the following relevant question: Does SISR phenomenon gives birth to any chimera patterns? if yes, what is the manifestation of the resulting chimera pattern?

In the present work we investigate the chimera patterns that arise out of the SISR phenomenon. For the study we consider the FitzHugh--Nagumo system, which is a paradigmatic model
of type-II excitability \cite{Semenov2023}. We report the occurrence of a new chimera pattern, namely self-induced stochastic resonance chimera (SISR-chimera) and discuss its unique features by characterizing it through quantitative measures. We show that unlike CR chimera, the SISR-chimera pattern strongly depends upon the ratio of time scale and noise intensity. Further, it is demonstrated that the SISR chimera can be found in a parameter zone far from the Hopf bifurcation threshold.

The rest of the paper is organized in the following way: The next section discusses the condition of getting SISR in a FitzHugh-Nagumo (FHN) model. Sec.~\ref{sec:net} describes the network model of coupled FitzHugh-Nagumo units in the presence of noise. In Sec.~\ref{sec:result} we explore the spatiotemporal behaviors including chimera patterns induced by SISR. The characterization of the chimera pattern in the parameter space is reported in this section. Finally, Sec.~\ref{sec:con} concludes the outcome of the whole study.

\section{Noisy FitzHugh--Nagumo system}  
We consider the following FitzHugh-Nagumo (FHN) model in the presence of Gaussian white noise \cite{sisr}:
\begin{eqnarray}
\epsilon\dot{x} &=& x-\frac{x^3}{3}-y+\delta_{1}\sqrt{\epsilon}{\xi(t)}, \nonumber\\
\dot{y} &=& x+a+\delta_{2}{\xi(t)},
\label{fhn}
	\end{eqnarray}
where $x$ and $y$ are the activator and inhibitor, respectively. $\epsilon>0$ is a small parameter that controls the time-scale separation between the activator and inhibitor. $a$ ($>0$) determines the excitability threshold: for $a>1$ the system is in the excitable zone, for $a<1$ the system shows a stable limit cycle, and at $a=1$ a Hopf bifurcation occurs. 
We consider Gaussian white noise, i.e.,$\langle\xi(t)\rangle=0$, and $\langle\xi(t)\xi(t')\rangle=\delta(t-t')$. 
$\delta_{1}$ and $\delta_{2}$ are positive constant. For $\delta_{1}=0$ and $\delta_{2}\ne 0$ one may have coherence resonance (CR) as reported by Pikovsky and Kurths \cite{fhn48}; on the other hand and if $\delta_{1}\ne 0$ and $\delta_{2}=0$ one have self-induced stochastic resonance (SISR) which was shown in  \cite{sisr,sisr21}.

\begin{figure}
\centering
\includegraphics[width=0.6\textwidth]{"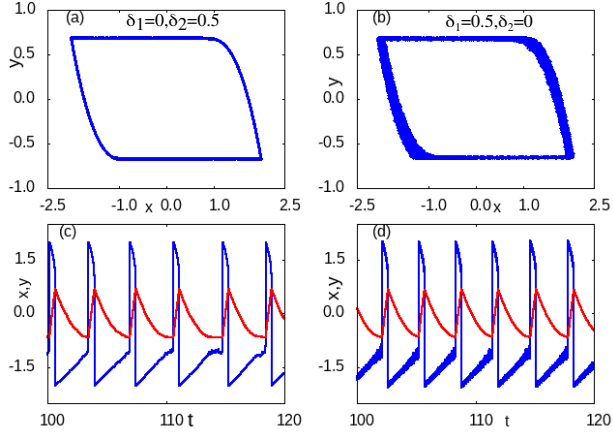"}
\caption{Limit cycle arises out of CR: (a) Phase-space, (c) time series for $\delta_{1}=0.0$, and $\delta_{2}=0.5$. Limit cycle arises out of SISR: (b) Phase-space, (d) time series for $\delta_{1}=0.5$, and $\delta_{2}=0.0$. In (c, d): $x$ in blue color and $y$ in red color. Parameter values: $\epsilon=0.005$, $a=1.001$.}
\label{fhn_a1.001}
\end{figure}

It may appear that the noise has a similar effect irrespective of whether it is added to the activator ($x$) or the inhibitor ($y$) or both. However, Ref.~\cite{sisr} demonstrated a marked distinction between
the two, particularly at smaller values of $\epsilon$. 
We demonstrate the phase plane and time series plots in Fig.~\ref{fhn_a1.001} for: (i) $\delta_{1}=0$, and $\delta_{2}=0.5$: showing the CR phenomenon and (ii) $\delta_{1}=0.5$, and $\delta_{2}=0.0$ showing the SISR phenomenon. 
We consider $a=1.001$ such that the deterministic model is in the excitable zone ($\epsilon=0.005$). Note the distinction between two phenomena: In the case of SISR the limit cycle is more noisy. Though, surprisingly, it was shown in \cite{sisr} that SISR is more ``coherent" than CR for the same set of noise intensity: the degree of coherence in SISR is
drastically increased and, moreover, the average period
$\langle T\rangle$ of oscillations shifted towards smaller values, which is by far greater than the uncertainty in $T$ \cite{sisr}.

%*****************************Section***2************************
%==========================begin network==========
\section{Network model}\label{sec:net}
Next, we consider a ring networks of $N$ identical excitable FitzHugh-Nagumo (FHN) systems, in the presence of noise, coupled through a non locally matrix coupling. The mathematical model of this network reads
\begin{eqnarray}
% \begin{split}
	\epsilon\dot{x}_{i} &= x_{i}-\frac{x_{i}^3}{3}-y_{i}+\frac{\sigma}{2P}\sum_{j=i-P}^{i+P}[b_{xx}(x_{j}-x_{i})\nonumber\\
	&\quad+b_{xy}(y_{j}-y_{i})]+\delta_{1} \sqrt{\epsilon} \xi_{i}(t), \\
	\dot{y}_{i} &= x_{i}+a_{i}+\frac{\sigma}{2P}\sum_{j=i-P}^{i+P}[b_{yx}(x_{j}-x_{i})\nonumber\\
	&\quad+b_{yy}(y_{j}-y_{i})]+\delta_{2}\xi_{i}(t).            
\label{fhn_net}
% \end{split}
	\end{eqnarray}
Here $i=1,2,3...N$, $\sigma>0$, is the coupling strength, coupling range $P\in[1,N/2]$ is the number of nearest neighbors of each oscillator on either side. The limit $P=1$ and $P=N/2$ give the nearest neighbor and global coupling, respectively. We have defined the coupling radius $r=P/N\in[1/N,0.5]$. $\delta_{1}$ and $\delta_{2}$ are the noise controlling parameter. We have used Gaussian white noise, i.e.,$\langle\xi_{i}(t)\rangle=0$ and $\langle\xi_{i}(t)\xi_{j}(t')\rangle=\delta_{ij}\delta{(t-t')}$. From Eq.~\ref{fhn_net} we have seen that not only the direct but also cross-coupling term is present between the activator($x_i$) and inhibitor ($y_i$) variables \cite{sniper25}, which is modeled by a rotational coupling matrix. The coefficient of $b_{lm}$, where $l,m \in [x,y]$, element of the rotational matrix:
$$
|B|=
\quad
\left(
\begin{array}{cc}
b_{xx}& b_{xy} \\
b_{yx} & b_{yy}\\
\end{array}\right)
=
\left(
\begin{array}{cc}
\cos{\phi}& \sin{\phi} \\
-\sin{\phi} & \cos{\phi}\\
\end{array}\right)
$$
where $\phi \in [-\pi,\pi]$. The matrix $\textbf{B} $ allows for direct coupling as well as cross coupling between $x$ and $y$ as in \cite{sniper25}. In this work we have used the parameter $\phi=\pi/2-0.1$ \cite{sniper25}, which acts as the phase lag parameter. 
%%%%
%================order parameter=================
To understand the spatial coherence and incoherence state of chimera pattern, we use the local order parameter \cite{{fhn15},{fhn79}} defined as
\begin{equation}
Z_{k}=\left\vert\frac{1}{2\delta_{m}}\sum_{|j-k|\leq \delta_{z}}e^{i\Theta_{j}}\right\vert,
\end{equation}
where $i=\sqrt{-1}$, $k=1,2,...N$, and $\delta_{z}$ is the nearest neighbors of the $j$-th node on the both sides. The geometric phase of the $j$-th element is defined by $\Theta_{j}=arctan(y_{j}/x_{j})$\cite{{sniper25}}. The local order parameter $Z_{k}\approx{1}$, denotes that the $j$-th oscillators is in the coherent domain of the chimera pattern. Again, if $Z_{k}<1$, it indicates that the $j$-th oscillators is in the incoherent domain of the chimera pattern \cite{{cr-chimera},{tbpre}}. In our work we take nearest neighbor $\delta_{m}=25$, and compute the local order parameter for a long time interval.

%=================results=======================
\section{Results}\label{sec:result}

In Ref.~\cite{cr-chimera}, Semenova et al. considered adding the noise in the inhibitor part only, i.e., they considered $\delta_{1}=0$ and $\delta_{2}\ne 0$ and observed coherence resonance chimeras. The same has been observed in type-I excitable systems \cite{type1}. However, to explore the SISR induced chimera patterns, {\it in the present work we only consider $\delta_{1}\ne 0$ and $\delta_{2}= 0$}. Throughout the study, unless stated otherwise, we take $a_{i}=a$ for all nodes, $a=1.001$ fixed at the excitable zone and very close to the Hopf bifurcation threshold. However, we also explored the SISR chimera patterns far from the Hopf bifurcation threshold. In the simulations random initial conditions distributed on a circle of radius $x_{i}^2+y_{i}^2=4$ are taken \cite{cr-chimera}.

%====================SISR===============fig
 \begin{figure}
\centering
\includegraphics[width=0.7\textwidth]{"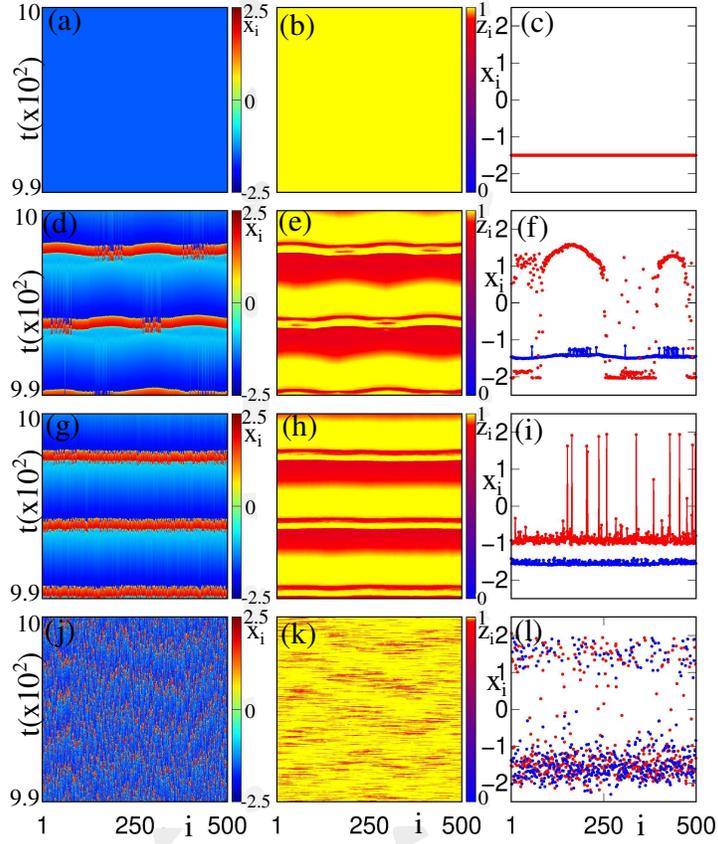"}
\caption{Space-time plots of ($x_{i}$) states (left column), Local order parameter($Z_{i}$) (second column, $\delta_{m}=25$) for different noise intensities. The corresponding snapshots are shown in the right column at two diferent time instances: $t=991$ (blue) and $t=993$ (red). (a,b,c): $\delta_1$=0.01: steady state, (d,e,f): $\delta_1=$ 0.045, self-induced stochastic resonance chimera, (g,h,i): $\delta_1=0.1$ incoherent in space and coherent in time. (j,k,l): $\delta_1$=0.5 complete incoherent pattern. Parameter values: $\epsilon=0.01$, $a=1.001$, $\delta_2$=0, $\sigma=0.1$, $P=60$ and $N=500$.}
\label{fhn_x}
\end{figure}

Depending on the noise intensity $\delta_{1}$, we investigate distinct spatiotemporal patterns arise in the network. For demonstration, we consider the following parameter values: $\epsilon=0.01$, $a=1.001$, $\delta_2$=0, $\sigma=0.1$, $P=60$ and $N=500$. We started with zero noise intensity ($\delta_{1}=0$), where all nodes in the network are in a homogeneous steady state (i.e., nonoscillatory state). This homogeneous steady state condition retains up to a certain limit of noise intensity $\delta_{1}\leq 0.035$. Figure~\ref{fhn_x}(a) shows this scenario for  $\delta_{1}\leq 0.01$ and the corresponding local order parameter is shown in Fig~\ref{fhn_x}(b). The snapshot of all the nodes at $t=991$ and $t=993$ are shown in Fig~\ref{fhn_x}(c); they coincide on each other indicating homogeneous steady state condition.
With increasing noise intensity, nodes show noise induced limit cycle ($\delta_{1}> 0.035$). We observe the self-induced stochastic resonance chimera (SISR-chimera) where a fraction of nodes are in the incoherent state and the remaining nodes are synchronized in spatial sense. Figure~\ref{fhn_x}(d) depicts the SISR-chimera for an exemplary value $\delta_{1}=0.045$; the corresponding local order parameter also supports the coexistence of coherent-incoherent states (Fig.~\ref{fhn_x}(e)). interestingly, the chimera pattern resembles a breathing type chimera and the coherent and incoherent zones switch their position with each cycle. The corresponding snapshot at $t=991$ (blue) and $t=993$ (red) are shown in Fig.~\ref{fhn_x}(f). 
With the increase of noise intensity, the incoherent zone spreads in space although the network remains coherent in temporal sense [see Fig~\ref{fhn_x}(g,h,i) for $\delta_1$=0.1]. Finally, at a strong noise intensity the network becomes incoherent in both space and time [see Fig~\ref{fhn_x}(j,k,l) for $\delta_1$=0.5].
 
 %===========snapshot_phase space==============

\begin{figure}
\centering
\includegraphics[width=0.6\textwidth]{"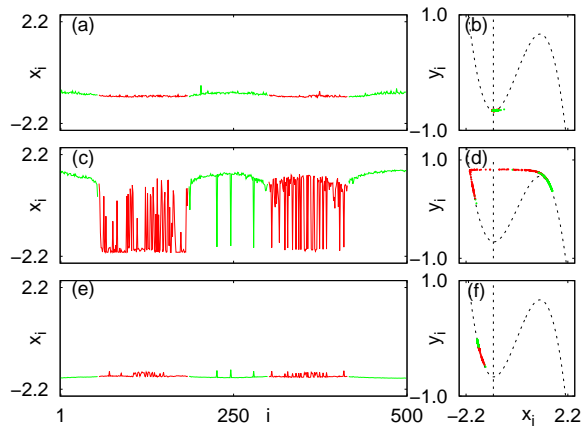"}
\caption{Snapshot and the corresponding phase-space: (a, b) $t=1981$, (c, d) $t= 1993$, (e, f) $t= 1995$. System parameters are $\epsilon=0.01$, $a=1.001$, $\sigma=0.1$, $P=60$ and $N=500$.}
\label{snapshot_fhn}
\end{figure}

The temporal evolution of activator $x_{i}$ and the corresponding phase plane plot in the $x_{i}-y_{i}$ plane are shown in Fig.~\ref{snapshot_fhn}. We show a snapshot at $t=1981$ when all the nodes of the networks are located near $x_{i}\approx-1$ [see Fig.~\ref{snapshot_fhn}(a,b)]. Fig.~\ref{snapshot_fhn}(b) shows that all the nodes lie near the left lower corner of the $x$-nullcline.  

Fig.~\ref{snapshot_fhn}(c,d) shows the SISR-chimera at $t=1993$ where the incoherent domains (shown in red) lies in between two coherent domains (shown in green). The phase plane plot [Fig.~\ref{snapshot_fhn}(d)] shows that the nodes corresponding to the incoherent domain (red points) are spreading along the phasespace, however, the nodes of the coherent domain are bunched near the right up corner of the $x$-nullcline. 
Again at $t=1995$ all the nodes come back to the lower branch [Fig.~\ref{snapshot_fhn}(e,f)] constituting the coherent domain. This scenario confirms the breathing nature of the SISR-chimera pattern.

%=======================impact of coupling parameters======
\subsection{Dynamical regimes}

\subsubsection{Impact of coupling parameters}
\begin{figure}
\centering
\includegraphics[width=0.5\textwidth]{"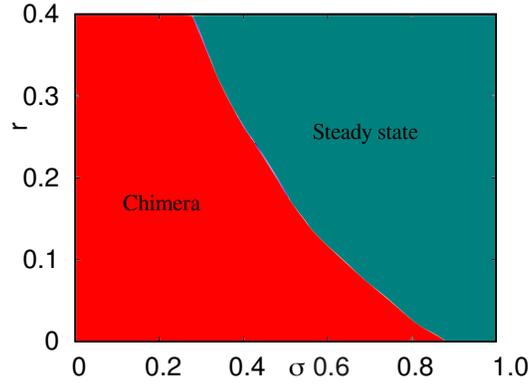"}
\caption{Dynamic regimes in ($\sigma-r$) parameter space. Incoherent zones are not shown as they occur in a very narrow region near $\sigma=0$. Parameters are $\epsilon=0.01$, $a=1.001$, $\delta_{1}=0.045$ and $N=500$.}
\label{twop}
\end{figure}

We map the distinct dynamical regimes of the network in the $\sigma-r$ parameter space. The following parameters are fixed at $\epsilon=0.01$, $a=1.001$, and $\delta_{1}=0.045$. Fig.~\ref{twop} shows that for a strong coupling $(\sigma)$ and a large coupling range ($r$) the network rests in a homogeneous steady-state. As $\sigma$ and $r$ decrease, a transition from homogeneous steady state to SISR-chimera occurs. Smaller coupling range and weak coupling strength are conducive for SISR-chimera. In the figure the zone of incoherence are not shown as they occur in a very narrow zone near the low coupling strength limit. 
%==================fig twop=====
%========== effects of timefactor and noise intensity===
\subsubsection{Impact of time scale}
\begin{figure}
\centering
\includegraphics[width=0.45\textwidth]{"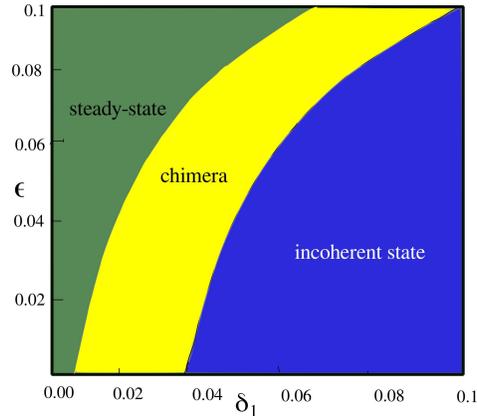"}
\caption{Dynamical regimes in $\delta_{1}-\epsilon$ parameter space. The fixed system parameters are $\sigma=0.1$, $a=1.001$, $r=0.12$ and $N=500$.}
\label{epsad}
\end{figure}

Next, we explore the effects of the interplay of time scale factor ($\epsilon$) and noise intensity ($\delta_1$) with fix network parameters. One of the characteristics of SISR is that it depends upon the ratio of the time scale factor and the noise intensity (unlike CR) \cite{sisr}. In our network due to the presence of a  perturbation in the activator section, we note that the network shows several distinct dynamical regimes depending on the time-scale and noise intensity. Fig.~\ref{epsad} presents the scenario in the two parameter space of $\delta_1-\epsilon$. Interestingly, we see that for a larger time-scale difference (i.e., smaller $\epsilon$) SISR-chimera appears even for a smaller noise intensity. Also, the SISR-chimera resides in between the steady state and incoherent state. Therefore, unlike the CR chimera, one can control the occurrence of the SISR-chimera by tuning the ratio of noise intensity and time scale factor.

\begin{figure}
\centering
\includegraphics[width=0.6\textwidth]{"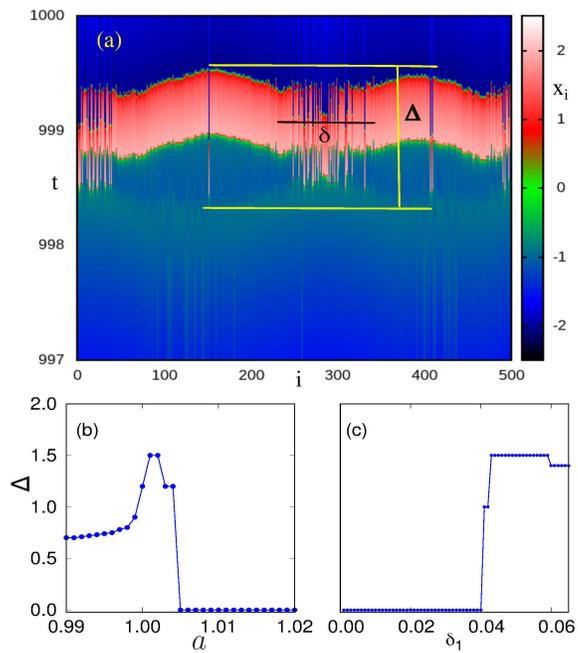"}
\caption{Characterization of chimera: (a) Space-time plot defining the span of active time $\Delta$ and incoherent domain $\delta$. (b) Variation of active time with excitable threshold $a$ ($\delta_1=0.045$), (c) Variation of active time with noise intensity $\delta_{1}$ ($a=1.001$). Other parameters are: $\sigma=0.6$, $P=60$ and $N=500$. }
\label{active_time}
\end{figure}
%===================active time span========
\subsection{Characterization of chimera pattern}
We further explore the impact of the excitable parameter and noise on the chimera pattern. 
Since spiking duration of the networks depends on excitable threshold $(a)$ and noise intensity ($\delta_{1}$), therefore, we will investigate their effect on the characteristic measure known as {\it active time span} of the chimera ($\Delta$), which means that the time separation between the first node belonging to the incoherent domain and return to the rest state [see Fig.~\ref{active_time}(a)]. Fig.~\ref{active_time}(b) and Fig.~\ref{active_time}(c) depict the variation of $\Delta$ on $a$ and $\delta_1$, respectively. Fig.~\ref{active_time}(b) shows that for a fixed noise intensity, the active time span increases with $a$ up to a certain excitable threshold. Beyond that, the network does not show any spiking behavior and suddenly collapses to the homogeneous steady state. Fig.~\ref{active_time}(c) shows that, with a fixed excitable threshold, with increasing noise intensity from zero the network at first rests in homogeneous steady state. When noise intensity $\delta_1\approx 0.045$ or above the network shows spiking nature, and  the active time-span increases from zero to a almost constant non zero value (although for a greater noise intensity the active time span duration decreases slightly). 

\begin{figure}
\centering
\includegraphics[width=0.8\textwidth]{"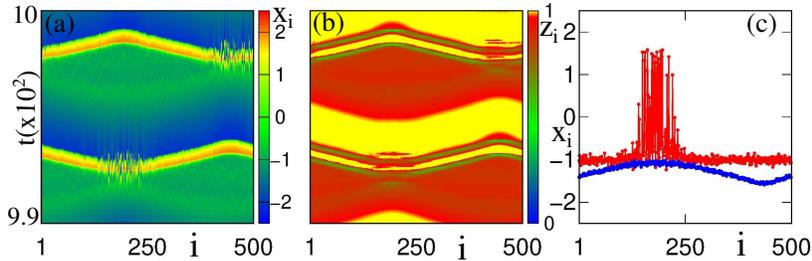"}
\caption{(a) Space-time plots of ($x_{i}$) states (left column), (b) Local order parameter ($Z_{i}$) (second column, $\delta_{m}=25$) (c) snapshots at $t=993$ (red) and $t=996$ (blue) for $\delta_1$=0.32. 
%different noise intensities (a,b): $\delta_1$=0: steady state,(c,d): $\delta_1$= 0.06,self induced stochastic resonance chimera, (e,f): $\delta_1$=0.1 incoherent in space but coherent in time.(g,h): $\delta_1$=5 complete incoherent. 
Parameter values: $\epsilon=0.05$, $a=1.02$, $\delta_2$=0, $\sigma=0.8$, $P=60$ and $N=500$.}
\label{fhn_a1.17}
\end{figure}
%%=======================very strong time factor======================
\subsection{Dyanmics of the system far away from excitable threshold}
\begin{figure}
\centering
\includegraphics[width=0.5\textwidth]{"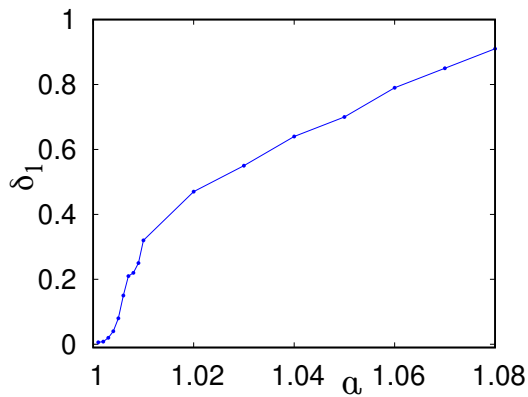"}
\caption{The excitability parameter ($a$) vs. the minimum noise intensity ($\delta_1$) required to bring SISR-chimera. Parameter values: $\epsilon=0.01$, $\delta_2$=0, $\sigma=0.6$, $P=60$ and $N=500$.}
\label{adelta}
\end{figure}

One of the prominent characteristics of the SISR phenomenon is that it occurs even at an excitation parameter far from the Hopf bifurcation threshold. SISR does not depend on the excitable threshold, i.e., it is not required that $a\rightarrow 1+$. In fact, we can choose the excitable parameter in any $1<a<\sqrt{3}$ \cite{sisr}.
In Fig.~\ref{fhn_a1.17} we demonstrate the SISR-chimera at $a=1.02$. It is seen that for $a$ values far from the threshold, the number of incoherent zone reduces to one. Fig.~\ref{fhn_a1.17}(a) shows the spatiotemporal plot of the SISR-chimera and the corresponding order parameter is shown in Fig.~\ref{fhn_a1.17}(b). The snapshots of the SISR-chimera for two different time instances [$t=993$ (red) and $t=996$ (blue)] are shown in Fig.~\ref{fhn_a1.17}(c). In our study, we noticed that the minimum noise intensity required to bring the SISR-chimera increases with increasing value of $a$. This is shown in Fig.~\ref{adelta}, which shows the variation of minimum noise intensity ($\delta_1$) required to bring the SISR-chimera ($\epsilon=0.01$). This can be attributed to the need of larger energy required to bring the system from an excitable state to the oscillatory state. 

 %=============conclusion===========================
 \section{Conclusion}\label{sec:con}
In this work, we have reported a new chimera pattern in a network of coupled excitable units. We called it self-induced stochastic resonance chimera or SISR-chimera because its genesis roots from the SISR phenomenon that is a distinct mechanism than coherence resonance (CR). We have explored the detailed dynamics of the SISR-chimera in the parameter space using proper quantitative measures. We have found that unlike CR chimera, the SISR-chimera pattern strongly depends upon the ratio of time scale and noise intensity. Therefore, this type of chimera pattern can be induced even for a tiny noise intensity if the time scale separation of the activator and inhibitor is large enough. We further found that the SISR chimera occurs in a zone far from the excitable threshold. 
The present study shows that the effect of noise in the excitable systems is not trivial and thus this study  deepens our understanding of neuronal networks in the presence of noise.\\
\noindent {\it Note added}: After the completion of this work we have come to know about the recent arXiv paper \cite{sisr-arxiv} that studied self-induced-stochastic-resonance breathing chimeras.
\section*{Acknowledgement} 
T. B. acknowledges the financial support from the Science and Engineering Research Board (SERB), Government of India, in the form of MATRICS Research Grant [MTR/2022/000179].



%  \bibliographystyle{elsarticle-num} 
%  \bibliography{references}


%\begin{thebibliography}{00}


%\bibitem{}

%\end{thebibliography}
\end{document}